\documentclass[]{spie}  

\usepackage{amsmath,amsfonts,amssymb}
\usepackage{graphicx}
\usepackage{dsfont}
\usepackage[colorlinks=true, allcolors=blue]{hyperref}

\title{Estimation-theoretic analysis of lensless imaging}

\author{Leyla A. Kabuli}
\author{Nalini M. Singh}
\author{Laura Waller}
\affil{Department of Electrical Engineering and Computer Sciences \\University of California, Berkeley, California, USA}

\authorinfo{Further author information: (Send correspondence to Leyla Kabuli)\\Leyla Kabuli.: E-mail: lakabuli@berkeley.edu}

\begin{document} 
\maketitle

\begin{abstract}
We analyze lensless imaging systems with estimation-theoretic techniques based on Fisher information. Our analysis evaluates multiple optical encoder designs on objects with varying sparsity, in the context of both Gaussian and Poisson noise models. Our simulations verify that lensless imaging system performance is object-dependent and highlight tradeoffs between encoder multiplexing and object sparsity, showing quantitatively that sparse objects tolerate higher levels of multiplexing than dense objects. Insights from our analysis promise to inform and improve optical encoder designs for lensless imaging.

\end{abstract}


\section{INTRODUCTION}
\label{sec:intro}  

Lensless imaging captures an image of an object using a non-traditional optical encoder, such as a phase mask placed near the sensor, rather than a lens placed further away. This captured image is a multiplexed measurement, in which each point of the object is mapped to multiple points in the measurement. The mask-based lensless design can be considered an optical encoder; when combined with a computational decoder, such as a deep learning-based reconstruction algorithm, we can recover an estimate of the object from the multiplexed measurement~\cite{diffusercam, Boominathan:22}. Lensless imagers are compact and low-cost imaging systems, providing a promising alternative to traditional lensed systems for applications including biomedical imaging and fluorescence microscopy.

To evaluate and improve designs for lensless imaging systems, we need a computational framework to characterize their performance.
 The traditional strategy for evaluating imager performance is to apply the reconstruction algorithm, then evaluate the reconstruction quality of the combined encoder-decoder system using quantitative metrics such as mean squared error. However, reconstruction-based evaluation cannot separate performance improvements due to the optical encoder from improvements due to the computational decoder. This dependence on the decoder, resulting in varying performance for different reconstruction algorithms and related regularization parameters, limits our ability to study and optimize the optical encoder in isolation. Our goal is to apply an alternate computational framework that characterizes the quality of optical encoders, regardless of the algorithm used to reconstruct their measurements.

To this end, we use tools from estimation theory that characterize the variance in estimating the underlying object, given the data from a particular optical encoder. Fisher information and the Cramér-Rao bound (CRB) provide a lower bound on this variance for an optimal unbiased estimator (i.e., optimal decoder or reconstruction algorithm). As this analysis assumes an optimal decoder instead of a specific implementation, this approach is a decoder-independent strategy for evaluating the optical encoder. These estimation-theoretic techniques have previously been used to guide the analysis and design of imaging systems for applications including single molecule localization microscopy~\cite{quirinOptimal3DSinglemolecule2012, shechtmanOptimalPointSpread2014b} and phase microscopy~\cite{bouchetFundamentalBoundsPrecision2021}.

In this work, we analyze lensless imaging using estimation-theoretic techniques. We model lensless imaging as a parameter estimation problem in a probabilistic framework, where our objective is to estimate the intensity values of our object. We analyze the performance of different phase mask encoders and consider two noise models: Gaussian noise, a simple model suitable for read noise-limited imaging conditions, and Poisson noise, suitable for shot noise-limited imaging conditions. By computing decoder-independent bounds on estimating object intensity values for each imaging condition, we find that bounds for Gaussian noise are object-independent and bounds for Poisson noise are lower for sparse objects than for dense objects. In general, bounds increase with encoder multiplexing and sparse objects tolerate higher multiplexing than dense objects.

\section{Analysis Framework}
\label{sec:framework}
\begin{figure}[t]
\includegraphics[width=\textwidth]{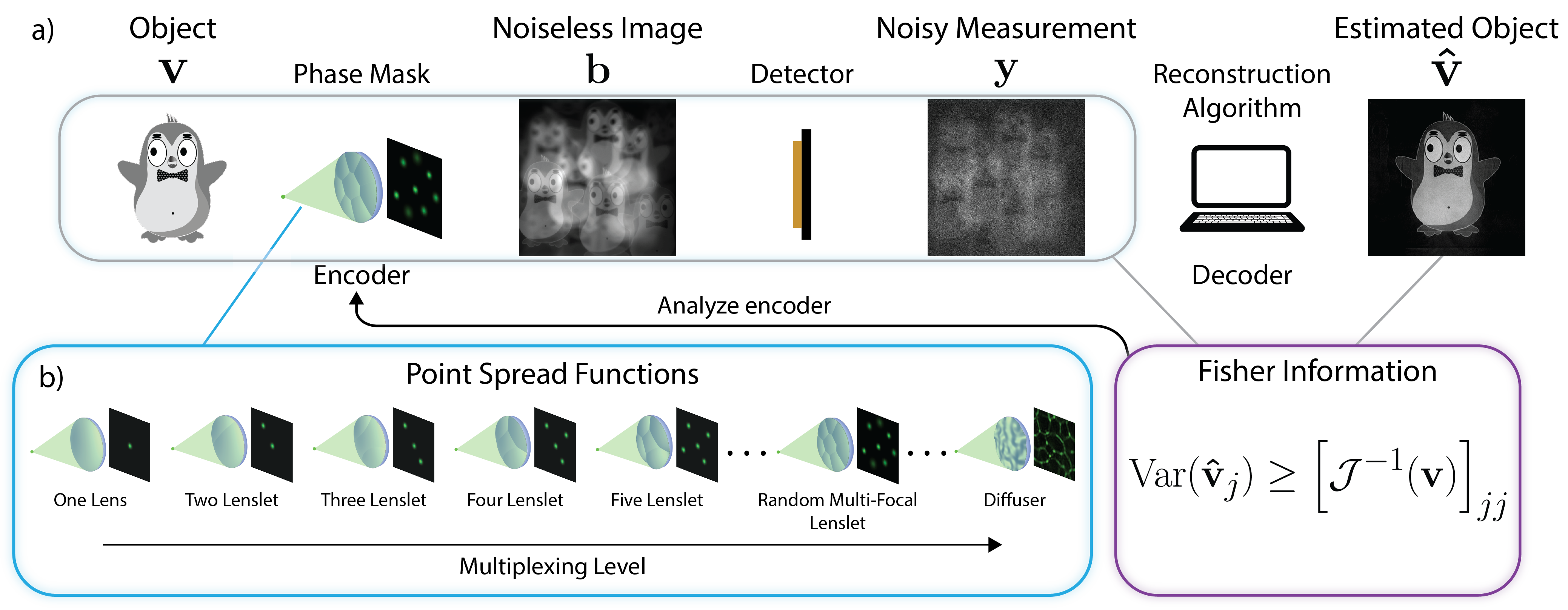}
\caption{\textbf{Lensless imaging:} \textbf{a)} Lensless imagers use a phase mask encoder to capture a multiplexed image of an object. The detector adds noise, forming a noisy measurement. A reconstruction algorithm decodes the measurement, providing an estimate of the object. The Fisher information can be used to analyze encoder performance by providing lower bounds on the variance of estimating each pixel $\mathbf{\hat{o}}_j$ of the object. \textbf{b)} We compare encoders with point spread functions with various levels of multiplexing, from a single lens to random multi-focal lenslet and diffuser phase mask encoders.}
\label{fig:intro}
\end{figure}

The image formation and object estimation model for lensless imaging is visualized in Fig.~\ref{fig:intro}a. 
The measurement $\mathbf{b}$ is modeled as a convolution between the object and the imaging system point spread function (PSF)~\cite{Boominathan:22}
\begin{equation}
    \mathbf{b} = \text{crop}(h \ast \mathbf{v}) = \mathrm{H} \mathbf{v}.
\end{equation}
Here $\mathbf{v}$ denotes the object being imaged through convolution with the system PSF $h$. The image formation model can also be expressed as a linear system with system matrix $\mathrm{H} \in \mathbb{R}_{+}^{k \times d}$ and the vectorized form of the object $\mathbf{v} \in \mathbb{R}_{+}^{d}$, where $\mathbf{v}$ has $d$ pixels and $\mathbb{R}_{+} = \{x \in \mathbb{R} : x \geq 0\}$. The measurement is cropped by the sensor extent. For simplicity, we omit the cropping operation in this analysis, equivalent to considering an object with finite extent and a sufficiently larger sensor. All objects and measurements in this analysis are represented on discretized pixel grids and take on real, non-negative values (intensities in photon counts per unit area).


The encoded measurement captured on a sensor includes detection noise, which introduces a source of randomness into the image formation model. This noise is typically modeled as Gaussian or Poisson (shot) noise that is independent and identically distributed (i.i.d.) at each pixel~\cite{goodmanStatisticalOptics2015}. For example, the noisy measurement $\mathbf{y} \in \mathbb{R}_{+}^{k}$ for additive Gaussian noise with noise vector $\mathbf{n} \sim \mathcal{N}(0, \Sigma)$, where $\mathbf{n} \in \mathbb{R}^{k}$ and $\Sigma \in \mathbb{R}^{k \times k}$, is 
\begin{equation}
    \mathbf{y} = \mathrm{H} \mathbf{v} + \mathbf{n}.
\label{eq:additivenoise}
\end{equation}

Given noisy measurement $\mathbf{y}$, the computational decoder aims to recover $\mathbf{v}$ by providing an estimate $\mathbf{\hat{v}}$ that is the solution to a least squares optimization problem with a non-negativity (intensity) constraint 
\begin{equation}
 \mathbf{\hat{v}} =\text{arg\,}\underset{\mathbf{v} \geq 0}{\text{min}} ||\mathbf{y} - \mathrm{H}\mathbf{v}||_2^2.
\label{optimization}
\end{equation}
An additional bias, a regularization term enforcing object sparsity, is often included in Eq.~\eqref{optimization}~\cite{Boominathan:22, diffusercam}.
As the objective of the decoder is to correctly estimate the intensity values of the object $\mathbf{v}$, the decoding process can be considered to be a parameter estimation problem where each parameter is a pixel intensity value~\cite{bouchetFundamentalBoundsPrecision2021}.

\subsection{Fisher Information}
The Fisher information, which relates a random variable $\mathbf{y}$ to the parameter set $\mathbf{v}$, is defined as
the outer product of the gradient of the log likelihood of the probability density function  $p(\mathbf{y}; \mathbf{v})$~\cite{Chao:16},
\begin{equation}
    \mathcal{J}(\mathbf{v}) 
     = \mathbb{E} \left[ 
     \left(\nabla_\mathbf{v} \ln 
     p(\mathbf{y};\mathbf{v})
     \right)
     \left( 
     \nabla_\mathbf{v} \ln 
     p(\mathbf{y};\mathbf{v})^\top
     \right)
     \right].
\label{eq:fisherinformation}
\end{equation}

This matrix $\mathcal{J}(\mathbf{v}) \in \mathbb{R}^{d \times d}$ quantifies the amount of information the random variable carries about the parameter set, specifically the sensitivity of the log likelihood to changes in each parameter. A high value for Fisher information is desirable, as it means that the random variable is sensitive to changes in the parameter. Increased sensitivity means higher precision in estimating the parameter value. A low value means that the random variable is not sensitive to changes in the parameter, leading to uncertainty and reduced precision for parameter value estimates.
The inverse of the Fisher information matrix is used to lower bound estimator performance with the Cramér-Rao bound~\cite{vantreesDetectionEstimationModulation2001}, such that the variance of any unbiased estimator satisfies  
\begin{equation}
    \text{Var}(\mathbf{\hat{v}}_j) \geq \left[ \mathcal{J}^{-1}(\mathbf{v}) \right]_{jj}.
\label{eq:crbequation}
\end{equation}
Higher Fisher information and lower variance are preferable for parameter estimation.

In order to compute the Fisher information matrix and resulting CRB for lensless imaging systems, we must model the probability distribution $p(\mathbf{y}; \mathbf{v})$. The convolutional forward model encoding the object into a noiseless image is a deterministic process, so this distribution captures only the randomness from the detection process.

\subsubsection{Gaussian noise model}

Gaussian noise follows the additive noise model in Eq.~\eqref{eq:additivenoise}. For i.i.d noise with per-pixel variance $\sigma^2$ and diagonal covariance matrix $\Sigma$, the probability density function is a normal distribution with mean vector equal to the noiseless image,
\begin{equation}
    p(\mathbf{y};\mathbf{v}) 
    = \mathcal{N}(\mathbf{b}, \Sigma)
    = \mathcal{N}(\mathrm{H}\mathbf{v}, \Sigma)
    = \frac{
    \exp(-\frac{1}{2} (\mathbf{y} - \mathrm{H}\mathbf{v})^\top 
    \Sigma^{-1}
    (\mathbf{y} - \mathrm{H}\mathbf{v})}
    {\sqrt{(2\pi)^k |\Sigma|}}
    .
    \label{eq:gaussiandist}
\end{equation}
Inserting Eq.~\eqref{eq:gaussiandist} into Eq.~\eqref{eq:fisherinformation}, the Fisher information matrix for Gaussian noise $\mathcal{J}_G$ has the form 
\begin{equation}
    \mathcal{J}_G(\mathbf{v}) = \mathrm{H}^\top \Sigma^{-1} \mathrm{H}.
\label{eq:gaussianfimatrix}
\end{equation}
Note that the Fisher information matrix has no dependence on the object, only the image formation model and the noise variance.

\subsubsection{Poisson noise model}

For Poisson noise, the relation between the random variable (measurement) and the object pixel intensities follows a Poisson distribution. For i.i.d. noise, the values at each pixel are statistically independent. The probability density function is then a Poisson random variable with rate parameter equal to the noiseless image~\cite{Chao:16, bouchetFundamentalBoundsPrecision2021},
\begin{equation}
    p(\mathbf{y}; \mathbf{v}) 
    = \prod_{\ell = 1}^k 
\frac{\mathbf{b}_{\ell}^{\mathbf{y}_{\ell}} e^{-\mathbf{b}_{\ell}}}
    {\mathbf{y}_{\ell} !}
    = \prod_{\ell = 1}^k 
\frac{(\mathrm{H}\mathbf{v})_{\ell}^{\mathbf{y}_{\ell}} e^{-(\mathrm{H}\mathbf{v})_{\ell}}}
    {\mathbf{y}_{\ell} !}
    .
\label{eq:poissondist}
\end{equation}
Inserting Eq.~\eqref{eq:poissondist} into Eq.~\eqref{eq:fisherinformation}, the Fisher information matrix for Poisson noise $\mathcal{J}_P$ has the form 
\begin{equation}
    \mathcal{J}_P(\mathbf{v}) = \mathrm{H}^\top \text{diag}((\mathrm{H}\mathbf{v})^{-1}) \mathrm{H},
\label{eq:poissonfimatrix}
\end{equation}
where $\text{diag}((\mathrm{H}\mathbf{v})^{-1})$ is a diagonal matrix with $i$th diagonal entry equal to $1 / (\mathrm{H}\mathbf{v})_i$. Appendix A provides full derivations of Eq.~\eqref{eq:gaussianfimatrix} and Eq.~\eqref{eq:poissonfimatrix}. Unlike the Gaussian noise case, the Fisher information matrix with Poisson noise depends on the object intensity values.

\section{Results}

We analyze lensless imaging systems in simulation using Fisher information and the CRB based on the framework described in Sec.~\ref{sec:framework}. We derive fundamental bounds on object intensity estimation from measurements encoded by various system matrices $\mathrm{H}$. Here, a better system matrix for lensless imaging is one that corresponds to lower bounds on the variance of estimating object intensities.

The system matrix $\mathrm{H}$ represents a multiplexing PSF, which maps each input point to many output points on the measurement. We systematically study multiplexing levels by considering PSFs with increasing number of lenslets, as well as two higher-multiplexing lensless imaging encoders: a random multi-focal lenslet phase mask~\cite{rml} and a diffuser phase mask~\cite{diffusercam}. These seven PSFs are visualized in Fig.~\ref{fig:intro}b.

All objects and PSFs are $32 \times 32$ pixels ($d = 1024$) and each system matrix $\mathrm{H}$ represents convolution including zero padding, such that measurements are $65 \times 65$ pixels ($k = 4225$). For each object and PSF we compute the Fisher information matrix, invert it, and extract the diagonals as described in Eq.~\eqref{eq:crbequation} to obtain the CRB. A small epsilon value is added along the diagonals of matrices before inversion for numerical stability.

\subsection{Gaussian Noise System}
We start by computing the Fisher information for each PSF under Gaussian noise, which corresponds to read noise-limited lensless imaging conditions. As shown in Eq.~\eqref{eq:gaussianfimatrix}, the Fisher information matrix does not depend on the object being imaged. Therefore, the quality of parameter estimation for systems with Gaussian noise will depend only on the encoding PSF and the noise level. Here we consider i.i.d. Gaussian noise with per-pixel variance $\sigma^2$ in photons, and resulting diagonal covariance matrix $\Sigma$.

\begin{figure}[ht]
\includegraphics[width=\textwidth]{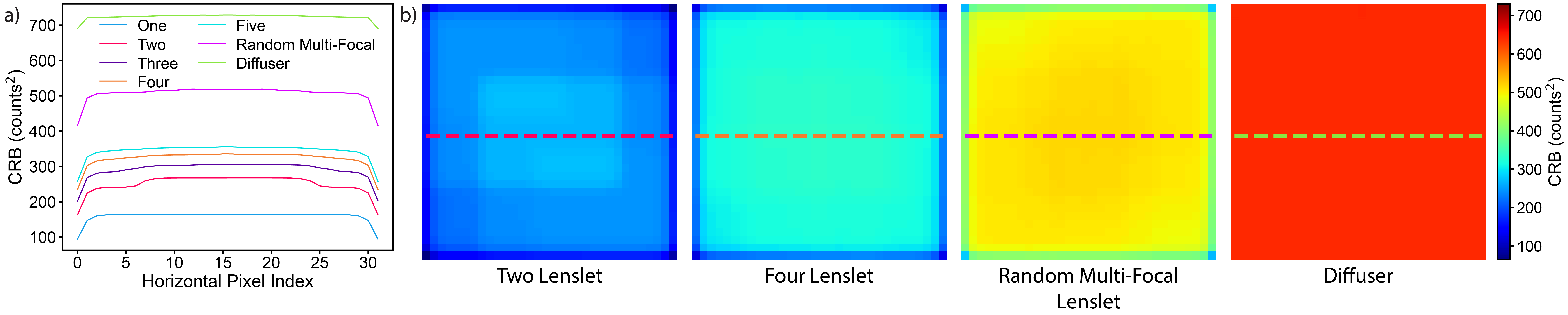}
\caption{\textbf{Parameter estimation with Gaussian noise:} The Cramér-Rao bound (CRB) for systems with additive Gaussian noise is independent of the object being imaged. \textbf{a)} As the encoder multiplexing level increases, the CRB gets worse, as visualized with a 32 pixel horizontal cross-section of the object region. \textbf{b)} Examples of the CRB for each pixel of the object region for different encoders. The horizontal cross-sections from $\textbf{a)}$ are indicated by dashed lines.}
\label{fig:gaussiancrb}
\end{figure}

Figure~\ref{fig:gaussiancrb} shows the resulting Cramér-Rao bounds on the variance of estimating each pixel of a $32 \times 32$ object region for each encoding PSF. Figure~\ref{fig:gaussiancrb}a plots the CRB for a horizontal cross-section through the center of the object region. As the multiplexing level of the encoder increases, the CRB gets worse, with the highest bounds on the variance for estimation with high-multiplexing systems such as the diffuser PSF. In Fig.~\ref{fig:gaussiancrb}b, the CRB for the $32 \times 32$ object region is visualized for various encoding PSFs. The two lenslet PSF case demonstrates more variance in estimating the center of the object, for which the measurement will have  multiple copies superimposed, and reduced variance towards the edges, where there is less overlap. This generalizes to increasing lenslets and saturates with the high-multiplexing diffuser PSF, where all of the points have significant overlap. This example with Gaussian noise provides insight into how increasing encoder multiplexing results in reduced Fisher information and higher bounds on the variance of estimating object pixel values.

\subsection{Poisson Noise System}
\label{sec:poissonfianalysis}
\begin{figure}[ht]
\includegraphics[width=\textwidth]{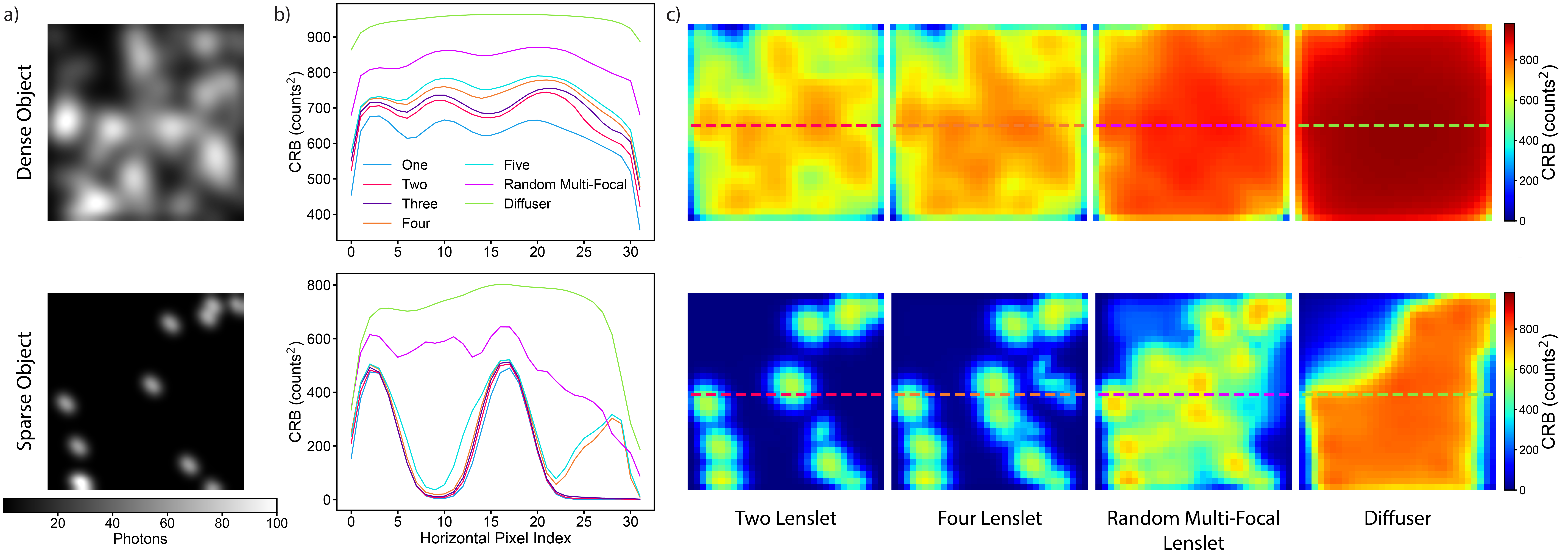}
\caption{\textbf{Parameter estimation with Poisson noise:} The Cramér-Rao bound (CRB) for systems with Poisson noise depends on the object being imaged. \textbf{a)} A dense sample of cell-like structures (top) and a sparse sample of beads (bottom). \textbf{b)} As the encoder multiplexing level increases, the CRB gets worse for the dense sample, as visualized with a 32 pixel horizontal cross-section of the object region. In contrast, for the sparse sample, the CRB remains low and approximately the same for one through five lenslets, increasing only for very high multiplexing levels. \textbf{c)} Examples of the CRB for each pixel of the object region for different encoders, where the CRB structure resembles the object being imaged in \textbf{a)}. The horizontal cross-sections from \textbf{b)} are indicated by dashed lines.}
\label{fig:poissoncrb}
\end{figure}

Next, we compute the Fisher information for each PSF under Poisson noise, which corresponds to shot noise-limited imaging conditions. As in Eq.~\eqref{eq:poissonfimatrix}, the Fisher information matrix depends on the object being imaged. To study this object dependence, we consider objects with varying degrees of spatial sparsity, where sparsity is quantified by the number of non-zero intensity values in the object. We consider a dense sample of cell-like structures and a sparse sample of beads, as visualized in Fig~\ref{fig:poissoncrb}a. Each object has maximum photon count of 100 photons.

Figure~\ref{fig:poissoncrb} displays the CRB for each object and encoder combination. For the dense object, the trend is similar to that for Gaussian noise in Fig.~\ref{fig:gaussiancrb}: as multiplexing levels increase, the bound on the variance of pixel estimates gets worse. For the sparse object, the CRB remains similar as multiplexing increases, with approximately equal CRBs for one through five lenslets, as visualized in Fig.~\ref{fig:poissoncrb}b. This is because there are fewer overlaps of the sparse object in a measurement. The CRB still gets worse at higher multiplexing levels with the random multi-focal lenslet PSF and the diffuser PSF. The CRB for each encoder with the sparse object is also lower than that for the respective encoder with the dense object. Overall, the object dependence of the CRB with Poisson noise is clear, as visualized in Fig.~\ref{fig:poissoncrb}c, where the CRB structure resembles the object being imaged.

\section{Discussion and Conclusion}

In this work, we analyzed lensless imaging systems using Fisher information and the Cramér-Rao bound. We calculated lower bounds on the variance of estimating  object intensities under Gaussian and Poisson noise models, different encoding PSFs, and objects with varying levels of sparsity. We demonstrated that systems with higher multiplexing levels have worse performance for all objects with Gaussian noise and for dense objects with Poisson noise. We discovered that sparse objects with Poisson noise are tolerant to higher levels of mutiplexing, maintaining low bounds on variance even as multiplexing levels increased.
To our knowledge, this is the first application of estimation-theoretic techniques to lensless imaging.

Our analysis is decoder-independent, focusing solely on the performance of the encoding half of lensless imaging.
The relative performance predicted by our analysis, where increased multiplexing encoders result in higher estimator variance, is in agreement with prior decoder-dependent reconstruction-based evaluations on dense objects~\cite{rml}, where higher multiplexing encoders have more reconstruction error.
Future work can compare the performance of different reconstruction algorithms to these lower bounds.  We note that the CRBs calculated in this work provide lower bounds that hold only for unbiased estimators, and that state-of-the-art reconstruction algorithms used in lensless imaging, such as deep learning algorithms, are rarely unbiased~\cite{Boominathan:22}. As dictated by the bias-variance tradeoff~\cite{hastieElementsStatisticalLearning2009}, biased reconstruction algorithms can have lower variance, exceeding the performance bounds of the unbiased CRB. Extensions of our analysis to the biased CRB or van Trees inequality would address this limitation~\cite{vantreesDetectionEstimationModulation2001}.

Our contribution is a decoder-independent approach for evaluating and improving encoders for lensless imaging.
Future work can use CRB-based design methodologies~\cite{shechtmanOptimalPointSpread2014b} to design phase masks for lensless imaging and extend these studies to higher-resolution objects and systems. In addition, combining our estimation-theoretic analysis with information-theoretic techniques~\cite{Kabuli:24, pinkard2024informationdrivendesignimagingsystems} and traditional reconstruction-based approaches for system evaluation and design can provide a complementary set of tools for improving lensless imaging systems.

\appendix    
\section{Derivations}
\label{sec:derivations}
In this appendix we derive the Fisher matrices used in the main text.
The Fisher information matrix in Eq.~\eqref{eq:fisherinformation} can be equivalently expressed as 
\begin{equation}
\mathcal{J}(\mathbf{v}) = 
    - \mathbb{E}
    \left[ 
    \nabla_{\mathbf{v}}^2
    \ln p(\mathbf{y}, \mathbf{v})
    \right].
\label{eq:fisheralternateexpression}
\end{equation}

\subsection{Gaussian noise model}

Following Eq.~\eqref{eq:fisheralternateexpression} with the probability distribution in Eq.~\eqref{eq:gaussiandist}, the log likelihood function is 
\begin{equation*}
    \ln p(\mathbf{y}, \mathbf{v}) = 
    -\frac{1}{2}
    (\mathbf{y} - \mathrm{H}\mathbf{v})^\top 
    \Sigma^{-1}
    (\mathbf{y} - \mathrm{H}\mathbf{v})
    -\frac{1}{2}
    \ln
    \left[ 
    (2 \pi)^k |\Sigma|
    \right].
\end{equation*}

The gradient $\nabla_{\mathbf{v}}$ is 
\begin{equation*}
    \nabla_{\mathbf{v}} \ln p(\mathbf{y}, \mathbf{v}) = \mathrm{H}^\top \Sigma^{-1}(\mathbf{y} - \mathrm{H}\mathbf{v}).
\end{equation*}

The Hessian $\nabla^2_{\mathbf{v}}$ is 
\begin{equation*}
    \nabla_{\mathbf{v}}^2 = 
    - \mathrm{H}^\top \Sigma^{-1}\mathrm{H}.
\end{equation*}

Therefore, as in Eq.~\eqref{eq:gaussianfimatrix},
\begin{equation*}
 - \mathbb{E}
    \left[ 
    \nabla_{\mathbf{v}}^2
    \ln p(\mathbf{y}, \mathbf{v})
    \right] = 
    \mathrm{H}^\top \Sigma^{-1}\mathrm{H}.
\end{equation*}

\subsection{Poisson noise model}

Following Eq.~\eqref{eq:fisheralternateexpression} with the probability distribution in Eq.~\eqref{eq:poissondist}, the log likelihood function is
\begin{equation*}
    \ln p(\mathbf{y}, \mathbf{v}) = 
    \sum_{\ell=1}^{k} 
    \left[ 
    \mathbf{y}_\ell \ln \mathbf{b}_\ell 
    - \mathbf{b}_\ell 
    - \ln(\mathbf{y}_\ell !)
    \right]
    = (\ln \mathbf{b})^\top \mathbf{y} - \mathbf{b}^\top \mathds{1}
    - \sum_{\ell=1}^{k} \ln (\mathbf{y}_\ell !),
\end{equation*}
where $\mathds{1} \in \mathbb{R}^{k}$ is a vector of all ones.

The gradient $\nabla_\mathbf{v}$ is 
\begin{equation*}
    \nabla_\mathbf{v} \ln p(\mathbf{y}, \mathbf{v}) = 
    \mathrm{H}^\top \text{diag}\left( (\mathrm{H}\mathbf{v})^{-1} 
    \right) \mathbf{y}
    = \mathrm{H}^\top 
    \left[ 
    (\mathrm{H}\mathbf{v})^{-1} \odot \mathbf{y}
    \right]
    ,
\end{equation*}
where $\odot$ denotes elementwise multiplication, $(\mathrm{H}\mathbf{v})^{-1}$ is a vector with $i$th entry equal to $1 / (\mathrm{H}\mathbf{v})_i$, and $\text{diag}((\mathrm{H}\mathbf{v})^{-1})$ is a diagonal matrix with $i$th diagonal entry equal to $1 / (\mathrm{H}\mathbf{v})_i$.

To compute the Hessian $\nabla_\mathbf{v}^2$, the matrix entries build on
\begin{equation*}
    \frac{\partial}{\partial \mathbf{v}_i}
    ((\mathrm{H}\mathbf{v})^{-1} \odot \mathbf{y}) = -\mathbf{y}_i (\mathrm{H}\mathbf{v})^{-2}_i \mathrm{H}_i,
\end{equation*}
where $\mathrm{H}_i$ is the $i$th row of $\mathrm{H}$. 
Generalizing this, 
\begin{equation*}
    \nabla_\mathbf{v}^2 \ln p(\mathbf{y}, \mathbf{v}) = \mathrm{H}^\top
    \text{diag}( - \mathbf{y} \odot (\mathrm{H}\mathbf{v})^{-2}) \mathrm{H},
\end{equation*}
where the $i$th entry of $(\mathrm{H}\mathbf{v})^{-2}$ is $1 / (\mathrm{H}\mathbf{v})_i^2$ .

Incorporating the expectation, where $\mathbb{E}[\mathbf{y}] = \mathrm{H}\mathbf{v}$, we arrive at Eq.~\eqref{eq:poissonfimatrix}.

\begin{equation*}
    - \mathbb{E}
    \left[ 
    \nabla_{\mathbf{v}}^2
    \ln p(\mathbf{y}, \mathbf{v})
    \right] = 
    - \mathrm{H}^\top \text{diag}( 
    - \mathbb{E}[\mathbf{y}] \odot (\mathrm{H}\mathbf{v})^{-2}) \mathrm{H} = \mathrm{H}^\top \text{diag}((\mathrm{H}\mathbf{v})^{-1}) \mathrm{H}
    .
\end{equation*}

\acknowledgments 
The authors thank Jonathan Dong and Henry Pinkard for helpful discussions.
L.A.K. was supported by the National Science Foundation Graduate Research Fellowship Program under Grant DGE 2146752. Any opinions, findings, and conclusions or recommendations expressed in this material are those of the author(s) and do not necessarily reflect the views of the National Science Foundation. N.M.S. was supported by the U.S. Air Force Office Multidisciplinary University Research Initiative (MURI) program under award no. FA9550-23-1-0281.

\bibliography{report} 

\begin{thebibliography}{10}

\bibitem{diffusercam}
Antipa, N., Kuo, G., Heckel, R., Mildenhall, B., Bostan, E., Ng, R., and Waller, L., ``Diffusercam: lensless single-exposure 3d imaging,'' {\em Optica}~{\bf 5}(1),  1--9 (2018).

\bibitem{Boominathan:22}
Boominathan, V., Robinson, J.~T., Waller, L., and Veeraraghavan, A., ``Recent advances in lensless imaging,'' {\em Optica}~{\bf 9}(1),  1--16 (2022).

\bibitem{quirinOptimal3DSinglemolecule2012}
Quirin, S., Pavani, S. R.~P., and Piestun, R., ``Optimal 3d single-molecule localization for superresolution microscopy with aberrations and engineered point spread functions,'' {\em Proceedings of the National Academy of Sciences}~{\bf 109}(3),  675--679 (2012).

\bibitem{shechtmanOptimalPointSpread2014b}
Shechtman, Y., Sahl, S.~J., Backer, A.~S., and Moerner, W., ``Optimal point spread function design for 3d imaging,'' {\em Physical Review Letters}~{\bf 113}(13),  133902 (2014).

\bibitem{bouchetFundamentalBoundsPrecision2021}
Bouchet, D., Dong, J., Maestre, D., and Juffmann, T., ``Fundamental bounds on the precision of classical phase microscopes,'' {\em Physical Review Applied}~{\bf 15}(2),  024047 (2021).

\bibitem{goodmanStatisticalOptics2015}
Goodman, J.~W.,  [{\em Statistical optics}{\nolinebreak\hspace{0.1em}]}, Wiley, New Jersey (2015).

\bibitem{Chao:16}
Chao, J., Ward, E.~S., and Ober, R.~J., ``Fisher information theory for parameter estimation in single molecule microscopy: tutorial,'' {\em J. Opt. Soc. Am. A}~{\bf 33}(7),  B36--B57 (2016).

\bibitem{vantreesDetectionEstimationModulation2001}
Van~Trees, H.~L.,  [{\em Detection, estimation, and modulation theory}{\nolinebreak\hspace{0.1em}]}, Wiley, New York (2001).

\bibitem{rml}
Kabuli, L., Wu, G., and Waller, L., ``High-quality lensless imaging with a random multi-focal lenslet phase mask,'' {\em Optica Imaging Congress} ,  CW3B.2, Optica Publishing Group (2023).

\bibitem{hastieElementsStatisticalLearning2009}
Hastie, T., Tibshirani, R., and Friedman, J.,  [{\em The Elements of Statistical Learning}{\nolinebreak\hspace{0.1em}]}, Springer Series in Statistics, Springer New York (2009).

\bibitem{Kabuli:24}
Kabuli, L.~A., Hung, C.~S., Markley, E., Pinkard, H., and Waller, L., ``Information-theoretic experimental analysis of lensless imagers,'' {\em Optica Imaging Congress} ,  CF1A.2 (2024).

\bibitem{pinkard2024informationdrivendesignimagingsystems}
Pinkard, H., Kabuli, L., Markley, E., Chien, T., Jiao, J., and Waller, L., ``Information-driven design of imaging systems,'' {\em arXiv} ,  2405.20559 (2024).

\end{thebibliography}
\bibliographystyle{spiebib} 

\end{document}